# Decarbonising price formation: unit-level evidence on battery storage and the imbalance price in the GB Balancing Mechanism

Robert Dalton, Aidan O'Sullivan

*UCL Energy Institute, University College London, Central House, 14 Upper Woburn Place, London WC1H 0NN, United Kingdom*

Corresponding author: Aidan O'Sullivan (aidan.osullivan@ucl.ac.uk)

## Highlights

- Marginal balancing actions are attributed to individual units, not fuel types
- Batteries set 36% of bid-side and 27% of offer-side margins by 2025 Q4
- Marginal capture per GW rose on both sides, outpacing fleet expansion
- Battery actions were about £10/MWh better than volume-matched substitutes
- Gas and pumped storage keep the price tail, so the GB paradox has moved there

## Abstract

Renewables now dominate Great Britain's generation mix but rarely occupy the marginal price-setting position, which raises the question of which flexible technologies translate a renewable-rich system into real-time price formation. This study reconstructs the price-ranked edge of the eligible bid or offer stack in the GB Balancing Mechanism across 50,684 half-hourly Settlement Periods from 2023 to 2025 and attributes it to individual Balancing Mechanism Units, separating long-system bid-active from short-system offer-active conditions and retaining co-marginal ties. Batteries rose from 0.8% to 36.2% of bid-active and from 3.2% to 26.9% of offer-active marginality, displacing combined-cycle gas on the bid side and pumped storage on the offer side. Capacity-normalised marginal capture rose on both sides, and unit-quarter fixed effects show that an additional 100 MW of Balancing Mechanism active capacity was associated with a 0.55 and 0.48 percentage-point increase in quarterly bid- and offer-side marginal share. By 2025, battery actions were around £9 to 10/MWh more favourable than volume-matched non-battery alternatives drawn from the same ladder, yet batteries were under-represented by 12.9 percentage points in the highest-priced 5% of short-system periods. Individual batteries remain consistent with price-taking, while the fleet has become endogenous to routine balancing price formation.

## Nomenclature

| Abbreviation | Definition |
|---|---|
| BESS | Battery energy storage system |
| BM | Balancing Mechanism |
| BMU | Balancing Mechanism Unit |
| BOD | Bid-Offer Data |
| CADL | Continuous Acceptance Duration Limit |
| CCGT | Combined-cycle gas turbine |

| Abbreviation | Definition |
|---|---|
| CfD | Contract for Difference |
| DISBSAD | Disaggregated Balancing Services Adjustment Data |
| FE | Fixed effects |
| HHI | Herfindahl-Hirschman Index |
| ISPStack | Settlement Bid-Offer Stack |
| NESO | National Energy System Operator |
| NIV | Net Imbalance Volume |
| OCGT | Open-cycle gas turbine |
| PAR | Price Average Reference |
| SO | System Operator |
| SP | Settlement Period |
| STOR | Short-Term Operating Reserve |
| VRES | Variable renewable energy sources |
| VWAP | Volume-weighted average price |
| S(i,t) | Share of total active-side marginal weight held by unit i in quarter t |
| K(i) | Registered power capacity of unit i (MW) |
| A(i,t) | Indicator equal to one where unit i was BM-active in quarter t |
| alpha(i) | Unit fixed effect |
| gamma(t) | Quarter fixed effect |
| beta | Coefficient on BM-active capacity |
| epsilon(i,t) | Error term |

## 1. Introduction

In liberalised electricity markets, short-run prices are set through marginal pricing: the final unit required to meet demand determines the market-clearing price [1,2]. Price levels alone, however, do not reveal which technologies set those prices or which respond to scarcity. Understanding the energy transition therefore requires identifying not only how prices change but which assets increasingly determine them.

Variable renewable energy sources (VRES) make this distinction more consequential. Wind and solar enter near the bottom of the merit order and suppress wholesale prices, while their variability raises the value of flexible resources close to delivery [3,4]. In Great Britain (GB), wind supplied 29.7% of generation in 2025 and renewables around 63%, while gas fell to 26.8% [5]. Yet in the sample analysed here, wind occupied the price-setting position in only 5% of Balancing Mechanism (BM) periods, against 37.1% for gas-fired technologies. Renewable forecast errors also create real-time imbalances, and annual GB balancing costs rose from £588 million in 2019 to £2.6 billion in 2022 [6]. Thermal price formation can therefore persist inside a decarbonising system, and gas-fired generation remains disproportionately important in European price setting during tight conditions [7].

These dynamics concentrate in the GB BM, through which the National Energy System Operator (NESO) resolves residual imbalances after gate closure. In each half-hourly Settlement Period (SP), NESO accepts bids and offers from Balancing Mechanism Units (BMUs). Bids absorb a long-system surplus by reducing generation or increasing demand, while offers resolve a short-system deficit by increasing generation or reducing demand. The two regimes are economically distinct, yet most empirical work pools them.

The most directly comparable GB study found that gas remained the marginal seller in 89.7% of identified periods between 2017 and 2022, with coal accounting for 4.2% and re-emerging as a high-priced seller of later resort, a pattern termed the British imbalance-market paradox [8]. That analysis was conducted at fuel-type level, largely pooled bid- and offer-side outcomes, and closed before the most recent phase of GB battery deployment.

Renewables may therefore reshape generation and balancing requirements without becoming dominant marginal price setters. Decarbonising price formation may instead depend on complementary flexible technologies that absorb renewable surpluses and respond during deficits. Battery energy storage systems (BESS) are the clearest candidate: by charging in long-system periods and discharging in short-system periods, they operate on both sides of the balancing problem. Falling lithium-ion costs have supported rapid deployment, and operational GB battery power capacity reached 6.5 GW by the end of 2025 [9,10]. Storage is nonetheless still modelled as a price-taking optimiser, with wider market influence examined through price compression or strategic market-power models [11,12,13]. Whether batteries have already become observed marginal price setters in a real balancing market has not been tested.

This study tests it directly, and finds that they have. Reconstructing the price-ranked edge of the eligible balancing stack across 50,684 half-hourly Settlement Periods from 2023 to 2025 and attributing it to named units, we find that batteries became routine occupants of the price-setting position on both sides of the market. Fig. 1 states the central result. Battery share of long-system bid-active marginality rose by 35.4 percentage points between the first and last quarters of the window, from 0.8% to 36.2%, and of short-system offer-active marginality by 23.7 points, from 3.2% to 26.9%. In both regimes this is by a wide margin the largest movement of any technology.

The two panels of Fig. 1 also show why the regimes must be analysed separately, because batteries advanced against different incumbents on each side. On the bid side the offsetting decline falls on combined-cycle gas, down 21.2 points, and wind, down 10.3 points, with pumped storage giving up only 5.9 points. On the offer side it falls almost entirely on pumped storage, down 22.9 points, while gas loses just 8.0 points and remains the leading short-system technology. The technology that batteries displaced when absorbing surplus energy is therefore not the technology they displaced when resolving deficits. A pooled ranking, which is how GB balancing-market price formation has previously been analysed, reports a single averaged decline for gas and shows neither pattern.

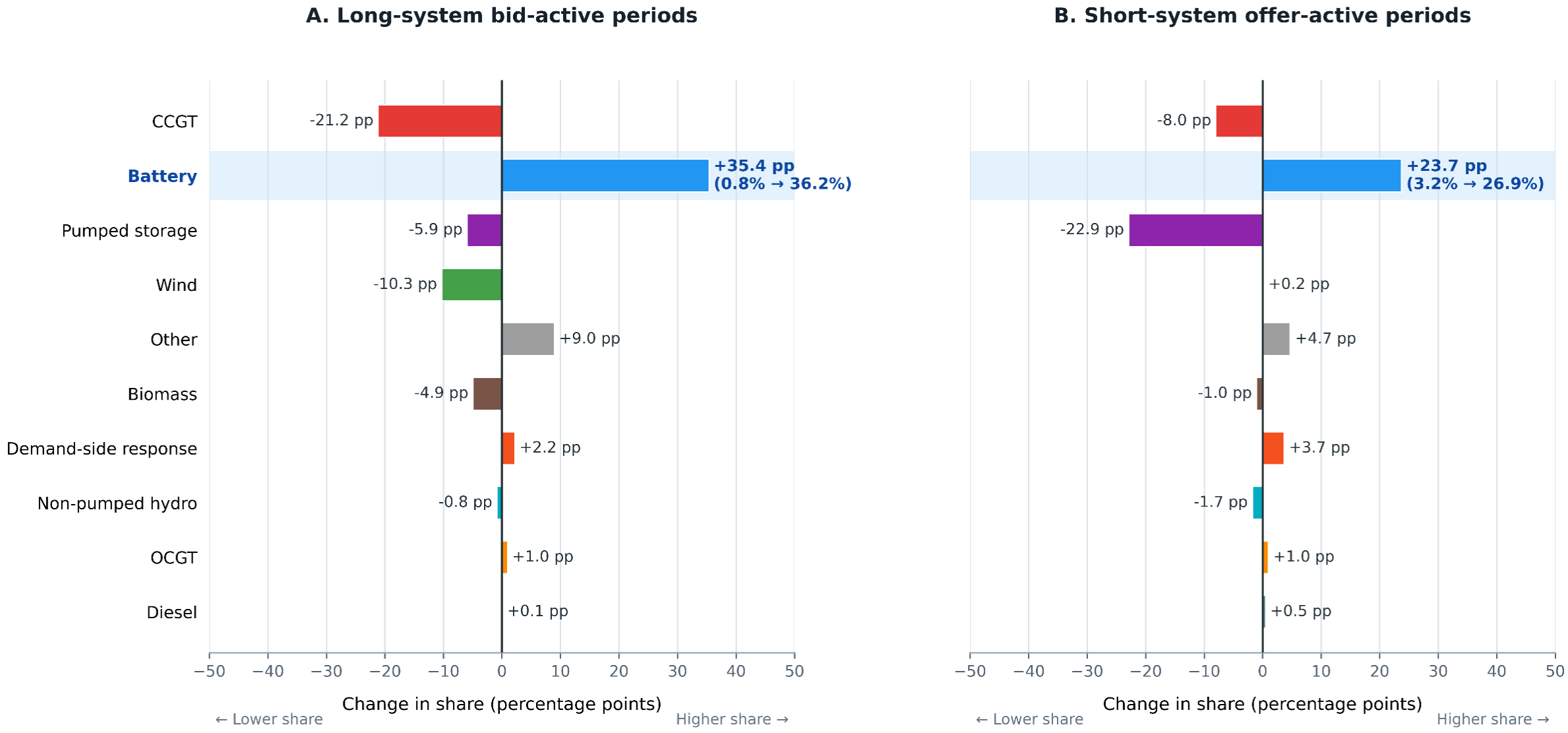

Fig. 1. Percentage-point change in technology shares at the price-setting position between Q1 2023 and Q4 2025, by active side. Panel A covers long-system bid-active periods and Panel B short-system offer-active periods. Shares are tie-adjusted within each active-side quarter. Batteries record much the largest gain in both regimes, but the offsetting incumbent decline is concentrated in combined-cycle gas and wind on the bid side and in pumped storage on the offer side.

Batteries did not, however, penetrate the upper price tail. Conditional on being marginal in a short-system period their prices sit close to those of gas and pumped storage at the median, yet the battery P95 reaches only £154.25/MWh against £192.00/MWh for gas and £203.00/MWh for pumped storage, and batteries are under-represented by 12.9 percentage points in the highest-priced 5% of short-system periods (Fig. 5). Battery price formation in the GB Balancing Mechanism is a phenomenon of the centre of the price distribution, not of its tail.

Three further results follow from the same reconstruction. *Marginal capture grew faster than the fleet*: capacity-normalised bid-active capture rose from 3.4 to more than 50 marginal periods per 1,000 active-side periods per GW, and unit-quarter fixed effects associate an additional 100 MW of Balancing Mechanism active capacity with 0.55 and 0.48 percentage points of quarterly bid- and offer-side marginal share. *Marginality dispersed rather than concentrated*: the number of marginal battery units rose from 65 to 137 while their leading-five share fell from 33.8% to 13.8%, and no individual battery reached the annual top 15 in any year of the window. *Accepted battery actions were better priced than the alternatives immediately behind them*: by 2025 the volume-matched non-battery replacement cost approximately £10/MWh more on both sides, a gap that widened rather than compressed as the fleet grew.

Together these results give the British imbalance-market paradox a new form. Fossil-fuelled technologies have lost much of their routine marginal role, but gas and pumped storage retain the highest-priced short-system outcomes, so the paradox has moved from routine marginality into the price tail.

The contribution is threefold. Methodologically, this is, to the authors' knowledge, the first ex-post study to reconstruct marginal balancing actions at named-unit resolution while separating long- and short-system regimes and preserving co-marginal ties, and to pair that reconstruction with a within-period fixed-ladder counterfactual built from submitted bid-offer pairs. Empirically, it establishes the frequency, dispersion and capacity association of battery marginality over a period of rapid deployment, and separates how often a technology occupies the price-setting position from where in the price distribution it does so. Conceptually, it shows that the price-taker and price-maker categories used throughout the storage literature are not exhaustive. No individual battery holds the share of the margin that unilateral price-making would require, yet the fleet collectively now sets the routine balancing price. Batteries occupy a third position that the existing taxonomy does not name, and one that should become more common wherever storage displaces fuel-based generation from the margin.

These findings are invisible at the resolution of existing work, which is why they have not previously been reported. Pooling bid- and offer-active periods conceals that batteries displaced gas on one side and pumped storage on the other. Fuel-type aggregation cannot say whether marginality rests on a few large assets or on a broad fleet, nor how unit size shapes that answer. And aggregate price-effect estimates cannot establish whether an accepted battery action was better priced than the alternative immediately available in the same period. Separating the regimes, resolving to named units and comparing within the period are what make these patterns legible.

Section 2 reviews the relevant literature. Section 3 sets out data and method. Section 4 reports results, Section 5 discusses their implications, and Section 6 concludes.

## 2. Background and related work

### 2.1 Marginal pricing under high renewable penetration

Under merit-order dispatch, generators are ordered by short-run marginal cost and the final unit required to meet demand becomes marginal; in uniform-price markets its offer sets the clearing price, which is intended to signal scarcity and support efficient dispatch and investment [1,2,14]. Liberalised markets were designed around dispatchable thermal plant with positive short-run marginal costs. Wind and solar, with high fixed and near-zero marginal costs, enter at the bottom of the merit order, displace higher-cost thermal generation and reduce residual demand, a mechanism commonly described as the merit-order effect [3,15,16].

VRES changes the composition of residual demand rather than the pricing mechanism itself. High renewable output lowers residual demand, increases the frequency of low and negative prices and reduces the marginal role of conventional thermal plant [3,4,17], while forecast uncertainty raises the value of assets able to respond close to delivery [4,18,19]. Ex-ante simulation established the merit-order effect as a theoretical expectation and linked VRES penetration to market-value decline and price cannibalisation [3,4,20,21,22]. A parallel ex-post literature confirms that higher renewable output lowers observed day-ahead prices [16,17,23], with the dampening effect strongest at upper price quantiles where gas-fired plant sets the price [24]. In both strands the marginal unit remains implicit in a cost stack or aggregated to national and technology level, rather than identified as a participant.

### 2.2 Balancing markets and the imbalance-market paradox

Price-formation research concentrates on day-ahead and, increasingly, intraday markets [25,26], with less attention to balancing despite its position between financial settlement and physical operation [27,28]. Early integration-cost work assumed that higher renewable penetration would raise balancing requirements [29,30,31]. The German balancing paradox challenged this view: despite rapid wind and solar growth, German balancing reserves fell rather than rose, a result attributed to improved forecasting, regional cooperation, market-design changes and greater intraday liquidity [32,33,34]. That literature redirected attention from how much balancing VRES requires towards how balancing is priced and which units set those prices [35,36].

Atherton et al. [8] apply this question to Britain and show that fossil-fuelled technologies, including coal, can retain or regain importance in imbalance-market price setting even as their role in the wider wholesale market declines. Three limitations remain. The analysis is conducted at fuel-type level and cannot identify which generators or balancing units set prices; accepted bids and offers are not fully separated despite their distinct roles under long and short system conditions; and the window closes before the rapid growth of battery participation after 2023.

## 2.3 Storage as price-taker, price-maker, or neither

BESS absorbs surplus generation, discharges rapidly during scarcity and provides short-duration flexibility [37], sitting at the intersection of declining VRES market value and rising flexibility value. Reviews map a storage literature spanning investment, market strategy, prices, models and security of supply [38], but storage is generally treated as an optimised flexibility asset rather than as an observed marginal price-setting unit.

Most battery research adopts the asset-owner perspective, initially modelling storage as an arbitrage resource [11,39,40] and later incorporating frequency response, degradation and multi-market revenue stacking [19,41,42]. European evidence suggests that energy arbitrage alone has often been insufficient, with stronger value in remunerated frequency services [12], and GB work optimises battery participation across frequency response and BM actions [13]. Storage is therefore usually modelled as a price-taker responding to an externally determined price path. A smaller literature asks whether storage affects prices directly, finding that charging in low-price and discharging in high-price periods compresses spreads and reduces peak prices and conventional generator revenues [43,44,45]. Alonso-Perez and Arcos-Vargas [46] find that 30 GWh of BESS roughly halves the average daily spread in Spanish day-ahead simulations, while Lamp and Samano [47] show that aggregate storage growth narrows intra-day spreads in California even where individual batteries remain price-takers. A further strand models storage as a strategic price-maker able to withhold capacity to preserve spreads, particularly under concentrated ownership [48,49,50,51,52,53]. Models of deeply decarbonised systems move closer to marginal price formation by treating storage as price-setting once fossil fuel leaves the merit order [54,55,56], but this evidence remains ex ante.

Observed storage price effects are strongest in ancillary services, where battery entry has compressed revenues, while GB balancing studies still focus on aggregate price drivers rather than individual marginal units [57,58,59]. The literature therefore establishes that storage affects price levels, spreads and welfare, and that it can become price-setting in modelled future systems, but not whether batteries already appear as observed marginal participants in a real balancing market.

## 2.4 Ex-post identification of marginal units

Observing price formation directly requires a method for identifying marginal units from market data. The dominant early approach inferred marginal shares from ex-ante fundamental models, identifying which technology set the price in each simulated hour [60,61,62,63]. Inferred shares depend on assumptions about demand, fuel costs, plant availability, interconnection and renewable output, and reflect modelled rather than observed dispatch, which is shaped by outages, forecast errors, transmission constraints and strategic bidding [7,64]. A more recent literature estimates marginal shares ex post from observed market data. Germeshausen and Wolfing [65] recover the marginal share of lignite in Germany from historical price and generation data, and Zakeri et al. [7] scale this approach across the EU-27, GB and Norway. These methods are transparent, reproducible and grounded in realised outcomes, but stop at national and technology-level resolution in day-ahead markets.

Three gaps therefore remain: pooled BM evidence obscures the difference between long-system bids and short-system offers; technology-level results cannot resolve concentration, unit size or unit-level capacity associations; and storage studies emphasise aggregate price effects rather than whether observed battery actions reach the margin at more favourable prices than nearby alternatives. This study addresses all three.

## 3. Data and method

### 3.1 Research design

The study uses an ex-post observational design covering 1 January 2023 to 31 December 2025. It reconstructs the active-side price-ranked stack after delivery and does not reproduce the dispatch decisions of NESO or estimate causal effects. Action-level settlement records are transformed into a long-form panel with one row for each distinct BMU attributed as marginal in an SP. Where several BMUs share the selected marginal price, each is retained as a separate co-marginal observation. Rows record BMU identity, technology, plant family, lead party, active side, selected action price, System Price and a tie weight.

### 3.2 Data sources and coverage

Settlement and bid-offer data were obtained through the Elexon Insights Solution API and supplemented with BMU reference information from Elexon and NESO. The study window contains 52,608 half-hourly SPs before filtering. The principal settlement input was the Settlement Bid-Offer Stack (ISPStack), which records the actions processed in the latest System Price calculation. It was preferred to raw operational acceptances because the study concerns settlement price formation. Bid-Offer Data (BOD) supply submitted price-volume pairs and unused volume behind each observed battery action, and are used only for the fixed-ladder exercise. A quarterly series of operational GB battery power capacity, measured in GW of operational power, contextualises fleet expansion and normalises aggregate marginal capture [10]. Table 1 summarises the inputs.

Table 1. Data inputs used in the analysis.

| Dataset | Provider | Resolution | Principal variables | Purpose |
|---|---|---|---|---|
| ISPStack | Elexon | Balancing action within SP | Asset identifier, bid or offer side, accepted volume, original and final price, action flags | Construct the eligible active-side stack and attribute the marginal action |
| System Price, NIV | Elexon | Settlement Period | Published System Price, Net Imbalance Volume | Identify the active side and attach the settlement price outcome |
| BMU reference metadata | Elexon, NESO | BMU or asset | BMU ID, technology, plant family, lead party, registered capacity | Resolve marginal actions to named assets and technologies |
| DISBSAD | Elexon | Balancing-service adjustment | Component identifier, asset, service classification, volume, cost | Resolve numeric ISPStack identifiers and retain energy-service actions |
| BOD | Elexon | BMU bid-offer pair within SP | Submitted bid and offer prices and volumes | Construct the fixed non-battery replacement ladder |
| Operational battery capacity | Modo Energy | Quarter | Operational GB battery power capacity (GW) | Contextualise fleet expansion and normalise aggregate marginal capture |

### 3.3 Sample construction and filtering

Eligible ISPStack rows were required to have a valid bid or offer direction, a non-null final price and non-zero accepted volume. Actions carrying Continuous Acceptance Duration Limit (CADL), System Operator (SO) or Short-Term Operating Reserve (STOR) flags were excluded, because these identify short-duration balancing, network or constraint management, or contracted reserve provision, and can receive distinct treatment in the System Price calculation. Their removal isolates a more comparable energy-balancing stack.

Some ISPStack rows use numeric component identifiers rather than named BMU identifiers. These were linked to Disaggregated Balancing Services Adjustment Data (DISBSAD) and retained only where the corresponding adjustment was classified as an energy service; all 1,326 such rows mapped to interconnectors. The filtered stack was then joined to System Price, Net Imbalance Volume (NIV) and BMU metadata by settlement date, SP and asset identifier. Positive NIV periods were classified as short and offer-active, negative NIV periods as long and bid-active. Zero-NIV periods, and periods without an eligible action on the active side, were excluded. The resulting analytical sample contains 50,684 SPs and 67,955 distinct marginal BMU-SP rows.

### 3.4 Side-resolved marginal attribution and co-marginal weighting

Marginal attribution was restricted to the NIV-active side in each SP. For short-system periods the selected marginal price was the maximum final offer price among eligible actions; for long-system periods it was the minimum final bid price. Every distinct BMU at that active-side price was retained, with repeated actions from the same BMU collapsed to one row per SP.

This procedure identifies the price-ranked edge of the eligible active-side energy stack. The Elexon Price Average Reference (PAR) volume is 1 MWh, so System Price is generally determined near the edge of the adjusted stack, and selecting the most extreme-priced eligible action is a practical marginal-action proxy. The procedure does not reproduce replacement pricing, de minimis or arbitrage removal, or the Reserve Scarcity Price floor, and is not a full Balancing and Settlement Code reconstruction.

The panel is stored in long form with System Price repeated across co-marginal rows. Each SP contributes a total weight of one, divided equally among its marginal BMUs, so that periods with several candidates do not receive greater influence. These tie weights are used in all shares, medians, concentration summaries and model estimation.

### 3.5 Unit identity and technology classification

BMU identity and technology were assigned hierarchically. Named ISPStack identifiers were matched to a curated BMU fuel-type register, with Elexon and NESO metadata as fallbacks. Plant families group sibling BMUs belonging to the same site or asset family. Generic gas labels were refined to combined-cycle gas turbine (CCGT) or open-cycle gas turbine (OCGT) where possible, interconnectors were assigned to named links, and a curated battery register overrode less specific labels. Unresolved technologies were retained as OTHER. All 771 marginal BMU identifiers were assigned a named unit, plant family and lead party; 59 units representing 4,918 rows (7.2%) remained in OTHER, principally aggregators or supplier portfolios with mixed assets.

### 3.6 Descriptive and temporal analysis

Technology and BMU marginal shares were calculated by summing tie weights within the relevant active side, year, quarter or NIV bin. Within each technology-year, concentration was summarised using marginal BMU counts, leading-five and leading-15 shares, and the Herfindahl-Hirschman Index (HHI), with lower values indicating broader participation. To test whether battery dispersion partly reflects smaller unit sizes, concentration was recalculated after dividing each BMU annual marginal weight by registered MW capacity. Comparing unadjusted and capacity-normalised concentration separates unit-scale effects from marginal participation per MW. BMUs without positive registered capacity were excluded from the normalised calculation, and the adjustment does not capture storage duration.

Battery marginal capture was separately normalised by fleet size as tie-adjusted battery-marginal periods per 1,000 active-side SPs per GW of operational capacity. NIV intensity was divided into eight fixed bins from below -1,000 MWh to above 1,000 MWh, and within each bin technology shares, median System Price and SP counts were calculated. Conditional System Price positioning was assessed using tie-weighted daily means, quarterly medians and annual distributions, with CCGT as the bid-side reference and CCGT and pumped storage as offer-side references. Upper-tail analysis compared technology representation in the highest 5% of offer-active System Prices with overall offer-side marginal shares.

## 3.7 Unit-level capacity association models

To examine the association between battery capacity and marginal participation, the analysis was resolved to battery BMU-quarter level. A complete panel of 142 battery BMUs across the 12 quarters from 2023 to 2025 yields 1,704 BMU-quarter observations. Quarterly bid- and offer-side marginal shares were calculated for each BMU by summing its tie-adjusted marginal weight and dividing by total active-side marginal weight in that quarter, so that BMUs not occupying the margin in a quarter receive a share of zero.

Observed BM activity was identified independently from the eligible ISPStack before marginal selection. A BMU was classified as BM-active in a quarter where it appeared in the eligible energy-action stack, and inactive otherwise. This measures observed participation rather than physical commissioning or continuous availability. Registered BMU power capacity was then interacted with activity status, and separate bid- and offer-side two-way fixed-effects models were estimated:

$$S_{i,t} = \beta(K_i \times A_{i,t}) + \alpha_i + \gamma_t + \varepsilon_{i,t} \tag{1}$$

where $i$ indexes battery BMUs, $t$ indexes quarters, $S$ is the unit share of total active-side marginal weight in that quarter, $K$ is registered power capacity in MW and $A$ is a binary indicator equal to one where the unit appeared in the eligible energy-action stack. The unit fixed effects absorb persistent differences between individual batteries and the quarter fixed effects absorb market conditions common to all batteries in a quarter, so the coefficient on the interaction captures the within-panel association between observed BM-active power capacity and unit marginal share. Standard errors were clustered by BMU.

A supporting specification adds a separate BM-active indicator to the same model, so that the association with appearing in the eligible stack at all is estimated alongside the capacity term rather than absorbed into it. This tests whether the capacity coefficient primarily reflects the transition from an unobserved to an observed BM-active quarter, or whether an additional size association remains once activity status is controlled separately.

## 3.8 Fixed-ladder replacement analysis

The fixed-ladder analysis combines ISPStack and BOD. ISPStack identifies each battery BMU attributed as marginal and the volume of its accepted action, while BOD supplies the submitted bid-offer pairs available in the same SP. BOD pair volumes were treated incrementally: volume already used in dispatch was removed before residual volume entered the replacement pool, after excluding battery and ineligible actions. For each battery-marginal observation the removed volume was refilled from unused non-battery actions on the same side, starting with the closest eligible price behind the battery action. Residual pairs were traversed once, using only the required share of the final action, and observations were retained only where the removed volume was fully replaceable.

The volume-weighted average price (VWAP) of the submitted battery action was compared with the volume-matched replacement VWAP. The offer gap is replacement minus battery VWAP; the bid gap is battery minus replacement VWAP. Positive values indicate a more favourable battery action, that is a lower offer or a higher, less negative bid. Rounded equality was classified as a tie, and co-marginal observations retained their SP-level weights. Because replacements are drawn from the next unused eligible positions behind the battery action, most comparisons are expected to favour the observed battery. The analysis is therefore interpreted through the magnitude and change in relative ladder distance rather than through the favourable sign alone. It does not reconstruct an unrestricted counterfactual dispatch, estimate consumer savings, or show that price alone determined operational selection.

## 4. Results

### 4.1 Marginal price formation recomposed asymmetrically across bid and offer regimes

Fig. 1 reported the net change in technology shares between the first and last quarters of the window. Fig. 2 traces the quarterly path behind it on each active side.

During long-system bid-active periods, battery share rose by 35.4 percentage points, from 0.8% to 36.2%, while CCGT and wind shares fell by 21.2 and 10.3 points respectively. Battery growth accelerated from 2024 and first exceeded CCGT in 2025 Q3. Batteries therefore became increasingly prominent in absorbing surplus energy, whether by increasing charging or reducing discharge.

During short-system offer-active periods, battery share increased by 23.7 percentage points, from 3.2% to 26.9%. This transition coincided primarily with a 22.9-point decline in pumped-storage share, while CCGT fell by only 8 points and remained the leading offer-side technology at 45% in 2025 Q4. Battery growth was therefore associated mainly with declining CCGT and wind participation on the bid side but with declining pumped-storage participation on the offer side. A pooled technology ranking would obscure both the pace of battery growth and the different incumbent structures against which it occurred.

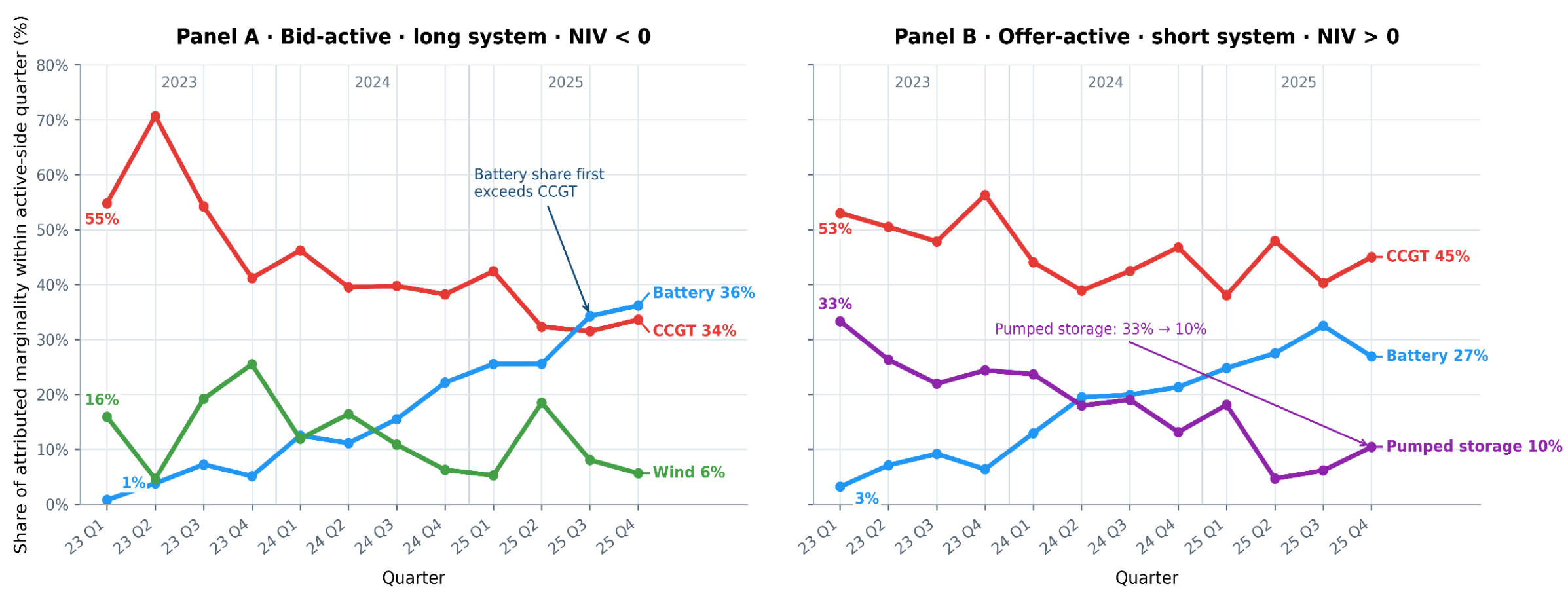

Fig. 2. Quarterly evolution of attributed marginality by active side, 2023 to 2025. Panel A shows long-system bid-active periods (NIV < 0) and Panel B short-system offer-active periods (NIV > 0). Shares are tie-adjusted within each active-side quarter.

## 4.2 Marginal capture grew faster than the fleet

The increase in battery marginality must be read against rapid fleet expansion. Operational battery power capacity rose from 2.45 GW in 2023 Q1 to 6.5 GW in 2025 Q4 (Fig. 3, Panel A). Fig. 3, Panel B therefore expresses tie-adjusted battery-marginal periods per GW of operational capacity per 1,000 active-side SPs.

Capacity-normalised marginal capture increased on both sides. Bid-active capture rose from 3.4 periods per 1,000 SPs per GW in 2023 Q1 to above 50 by late 2025, peaking at 55.6 in 2025 Q3. Offer-active capture rose from 13.3 to 41.4 and also exceeded 50 in 2025 Q3. Batteries therefore reached the price-setting position increasingly often relative to operational fleet power, so aggregate growth was not simply proportional to deployment.

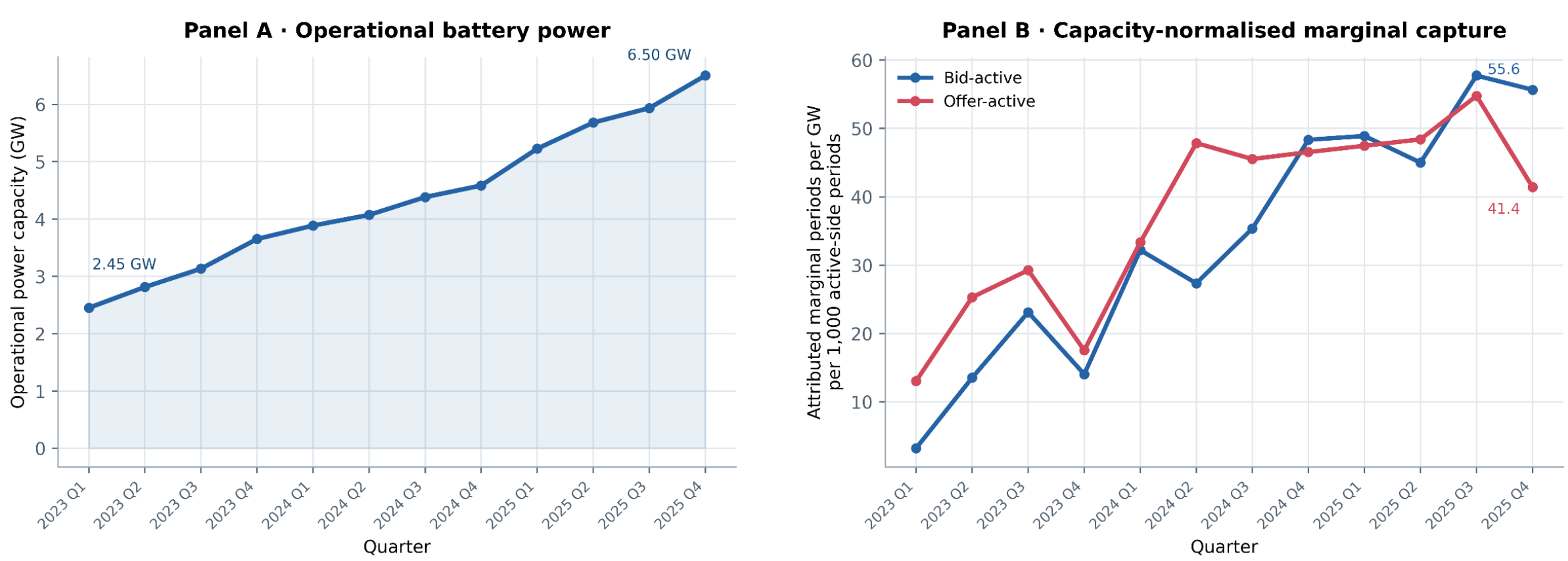

Fig. 3. Operational battery power capacity and capacity-normalised marginal capture by active side, 2023 to 2025. Panel B reports tie-adjusted battery-marginal periods per 1,000 active-side Settlement Periods per GW of operational capacity.

Disaggregating by the direction and intensity of NIV shows where in the space of system conditions the recomposition occurred. Across the pooled sample, wind dominated the deepest long-system conditions, accounting for 52% of marginal weight against 8.3% for CCGT, a concentration in the most negative-price conditions that is consistent with support-scheme effects on curtailment bids, since Contract for Difference generators may reflect foregone top-up payments in the prices submitted to reduce output [66]. Near zero NIV, wind fell to 7.8% and CCGT rose to 49%. CCGT retained the same 49% share immediately either side of zero, yet median System Price rose from £61/MWh to £110/MWh, so observed price outcomes varied with both the marginal technology and the active side. Battery share was comparatively stable across the long-system bins at 14% to 17% but rose from 16% to 25% as short-system imbalance intensified, while CCGT fell from 49% to 33% over the same range.

Comparing each bin between 2025 and 2023 shows that battery marginal share increased in all eight NIV bins, so its expansion was not confined to a single system condition. Long-system gains ranged from 19.8 percentage points in the deepest surplus bin to 28.3 points between -1,000 and -500 MWh, while short-system gains reached 29.8 points between 500 and 1,000 MWh and an estimated 30.1 points above 1,000 MWh. Both outer bins contain fewer than 150 periods in each year, so the consistent direction of change is more robust than the exact magnitudes. CCGT share fell in all eight bins and pumped-storage share in all four short-system bins, so CCGT retained aggregate offer-side leadership despite losing share throughout the short-system distribution. Full bin-level results are reported in the supplementary material.

## 4.3 Unit-level capacity associations, concentration and turnover

Technology-level results establish regime-specific recomposition but cannot show whether capacity is associated with marginal participation at individual-asset level, whether marginality is concentrated among a small group of units, or how stable the leading units are over time.

The BMU-quarter fixed-effects models show a positive association between BM-active capacity and marginal participation on both active sides (Table 2). In the primary specification, an additional 100 MW of BM-active capacity was associated with a 0.547 percentage-point increase in quarterly bid-side marginal share and a 0.484-point increase in offer-side marginal share, both significant at the 0.1% level, with within-R-squared values of 0.156 and 0.166. The relationship remained after separating the association with being BM-active from that of MW capacity, at 0.513 and 0.331 percentage points per 100 MW. The capacity association is therefore not explained solely by the transition between inactive and BM-active quarters. It was particularly stable on the bid side, while offer-side marginal participation reflected both BM-active status and unit size.

Table 2. BMU-level fixed-effects estimates of battery marginal participation. Coefficients are percentage points of quarterly active-side marginal share per 100 MW of BM-active capacity. Standard errors clustered by BMU; panel of 142 battery BMUs across 12 quarters (1,704 observations).

| Specification | Bid (pp per 100 MW) | p-value | Offer (pp per 100 MW) | p-value | Within $R^2$ bid | Within $R^2$ offer |
|---|---|---|---|---|---|---|
| Primary two-way FE | 0.547 | $< 0.001$ | 0.484 | $< 0.001$ | 0.156 | 0.166 |
| With separate BM-active control | 0.513 | $< 0.001$ | 0.331 | $< 0.001$ | 0.156 | 0.176 |

Annual rankings reveal persistence alongside turnover among recurrent marginal units. The leading BMU shifted from T_HUMR-1 in 2023 to T_SEAB-1 in 2024 and T_MRWD-1 in 2025, although all three were CCGTs, and the annual share of the leading unit fell from 4.02% to 3.02% and then 2.18%. The 15 largest units accounted for 30.6% of annual attributed marginal weight in 2023, 28.8% in 2024 and 20.4% in 2025, so repeated marginality became less concentrated among the leading group. Battery storage followed a distinct pattern: no battery entered the annual top 15 even as battery marginality rose sharply at technology level, and the highest-ranked battery moved only from 36th in 2023 to 29th in 2024 and 23rd in 2025. Leadership therefore remained with established non-battery units, particularly CCGTs, while battery prominence arose through participation across many individually less frequent units.

Fig. 4 and Table 3 show that unadjusted battery marginality became increasingly dispersed across individual BMUs, and that CCGT marginality did not. Between 2023 and 2025 the number of marginal battery BMUs rose from 65 to 137, the leading-five share fell from 33.8% to 13.8%, and HHI fell from 0.0481 to 0.0151. The shift is clearest at the top of the distribution: the leading battery held 17.6% of battery marginality in 2023 but only 2.9% in 2025, while units outside the leading 15 grew from 43.5% to 64.8% of the technology total. CCGT concentration over the same period was close to static, with 51 to 57 marginal units, the leading unit falling only from 7.5% to 5.7%, and units outside the leading 15 holding 47.5% and then 47.3%. The contrast in Fig. 4 is therefore between a technology whose marginal participation spread across an expanding fleet and one that retained a stable and recognisable core, and it is what makes the capacity-normalised comparison that follows necessary.

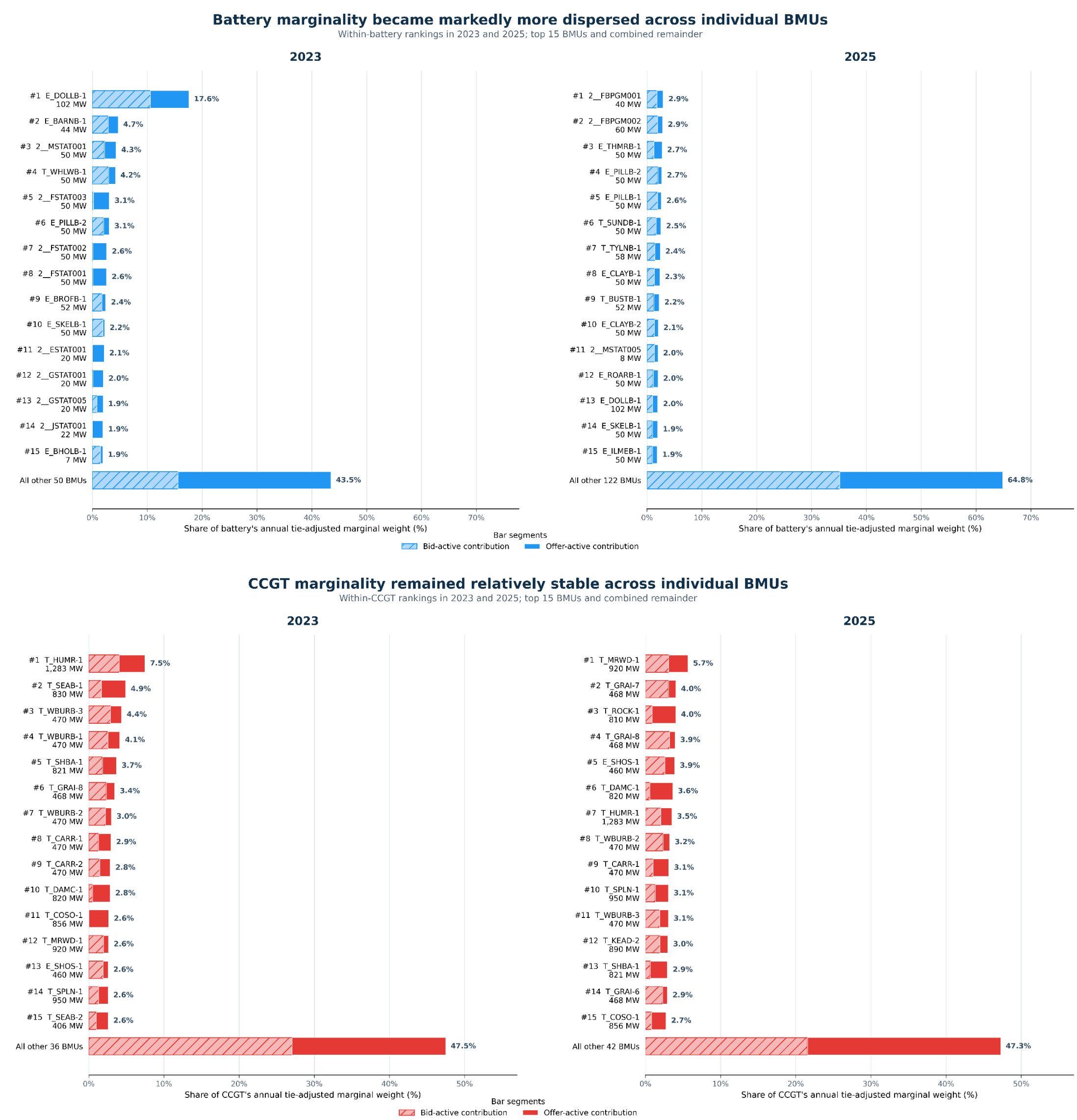

Fig. 4. Distribution of marginal weight across individual BMUs in 2023 and 2025, showing the 15 largest units and the combined remainder within each technology. Panel A covers battery storage and Panel B combined-cycle gas. Shares are within-technology, so each panel sums to the technology total.

Part of this contrast reflects differences in unit size, since battery BMUs are generally smaller than CCGTs and unadjusted marginal weight can make their fleet appear more fragmented. Capacity normalisation narrows the difference substantially. In 2025, battery HHI rose from 0.0151 unadjusted to 0.0234 after MW normalisation, against 0.0268 and 0.0256 for CCGT. Batteries therefore remained slightly more dispersed across the capacity-adjusted distribution, but the difference was modest. The leading-five comparison reversed: the five leading batteries accounted for 27.0% of capacity-normalised marginality against 20.9% for CCGTs, despite representing only 13.8% of unadjusted battery marginal weight. Battery marginality was therefore distributed across a broad and expanding fleet, but capacity normalisation reveals greater concentration among the most intensively marginal batteries per MW. These measures describe how often and how widely batteries reached the margin, but not the prices at which they did so, which Section 4.4 addresses.

Table 3. Unadjusted and capacity-normalised within-technology concentration of attributed marginality, 2023 to 2025.

| Year | Technology | Marginal BMUs | Unadjusted top 5 | Capacity-normalised top 5 | Unadjusted HHI | Capacity-normalised HHI |
|---|---|---|---|---|---|---|
| 2023 | Battery | 65 | 33.82% | 25.16% | 0.0481 | 0.0278 |
| 2024 | Battery | 96 | 16.13% | 25.25% | 0.0181 | 0.0240 |
| 2025 | Battery | 137 | 13.84% | 26.99% | 0.0151 | 0.0234 |
| 2023 | CCGT | 51 | 24.49% | 21.41% | 0.0282 | 0.0259 |
| 2024 | CCGT | 56 | 29.84% | 21.96% | 0.0340 | 0.0270 |
| 2025 | CCGT | 57 | 21.56% | 20.90% | 0.0268 | 0.0256 |

## 4.4 Battery price positioning at the centre and in the upper tail

Because the method retains every BMU tied at the marginal price, fleet dispersion could in principle be inflated by several batteries appearing in the same period. By 2025, however, most battery-marginal periods contained a single candidate, with co-marginal shares of 14.4% on the bid side and 10.5% on the offer side. Where ties did occur they were usually battery-clustered: in every year and on both sides, at least 74% of battery co-marginal sets contained multiple battery BMUs, rising in 2025 to 86.9% on the bid side and 77.1% on the offer side, with battery-only sets accounting for 78.4% and 69.0% respectively. Fleet dispersion is therefore not an artefact of tied prices.

Turning to price environments, bid-side daily mean System Prices were consistently lower when batteries were marginal than when CCGTs were, with the gap widening from £4.9/MWh in 2023 to £8.3/MWh in 2025. Offer-side averages were £3.0/MWh lower in 2023 and £3.2/MWh lower in 2025 but £2.1/MWh higher in 2024. The underlying distributions differed in shape as well as level. Bid-side battery-marginal periods showed greater density below £20/MWh and around zero, particularly in 2024 and 2025, whereas CCGT observations concentrated at higher prices. This is consistent with different underlying economics, since CCGT bids may partly reflect avoided fuel and operating costs whereas battery opportunity costs are not directly tied to contemporaneous gas prices. Offer-side distributions converged over time: in 2023 CCGT had a visibly heavier right tail around £170 to 220/MWh, while by 2025 the distributions were closely superimposed around the centre with only limited additional CCGT density above roughly £150/MWh.

Quarterly median gaps trace the same movement. On the bid side the battery-minus-CCGT gap moved from +£3.50/MWh in 2023 Q1 to -£5.02/MWh in 2025 Q4. On the offer side the battery-minus-CCGT gap narrowed from +£5.00/MWh to -£1.50/MWh, and the battery-minus-pumped-storage gap moved from +£1.00/MWh to -£6.45/MWh. The quarterly paths were not monotonic, but on both sides battery-marginal System Prices moved from a modest relative premium towards parity with, or below, those associated with the established comparators.

This convergence did not extend into the highest-priced short-system periods (Fig. 5). At the centre of the offer-side distribution, battery-marginal prices were broadly comparable with incumbents, with a battery median of £107.00/MWh against £113.90/MWh for CCGT and £119.00/MWh for pumped storage. The battery P95 nonetheless reached only £154.25/MWh, well below £192.00/MWh for CCGT and £203.00/MWh for pumped storage. Applying a common offer-side P95 threshold of £190/MWh, battery storage was under-represented by 12.9 percentage points relative to its overall offer-side marginal share, while CCGT was over-represented by 3.8 points and pumped storage by 12.8 points. Batteries had therefore become routine offer-side marginal participants without equivalent penetration of the highest-priced short-system outcomes.

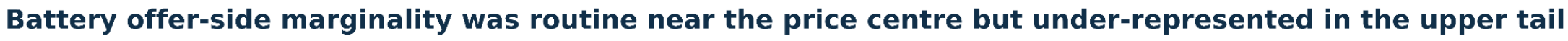

Fig. 5. Central and upper-tail offer-side System Price positioning by price-setting technology. Panel A reports the median and P95 System Price conditional on each technology being marginal; Panel B compares each technology share above a common £190/MWh threshold with its overall offer-side marginal share.

## 4.5 Within-period comparison against non-battery alternatives

The comparisons above draw on different Settlement Periods. The fixed-ladder exercise provides the stricter within-period test, comparing the submitted action-price VWAP of each marginal battery with a volume-matched non-battery replacement drawn from the next unused eligible actions on the same side of the observed ladder. Fully replaceable comparisons retained at least 99.8% of observations in every year-side cell.

Because replacement actions sit behind the observed battery on the ladder, most comparisons are expected to favour the battery, so the substantive result is how far that advantage widened (Fig. 6). Excluding batteries from the replacement pool, the mean gap rose from £5.64/MWh in 2023 to £10.24/MWh in 2025 on the bid side, and from £2.07/MWh to £10.23/MWh on the offer side. The share of comparisons favouring the battery rose from 71% to 91% on the bid side and from 66% to 93% on the offer side. Admitting other batteries to the replacement pool reduced the 2025 gaps only to £9.29/MWh and £9.20/MWh, so the widening did not depend on excluding battery substitutes. The result persisted after normalising the gap by both the prevailing System Price and the spread of the surrounding submitted ladder, indicating that the widening was not simply an artefact of higher prices or wider ladder spacing. Batteries therefore became both more frequent and more favourably positioned around the routine margin, while CCGT and pumped storage retained greater reach into the upper offer-price tail.

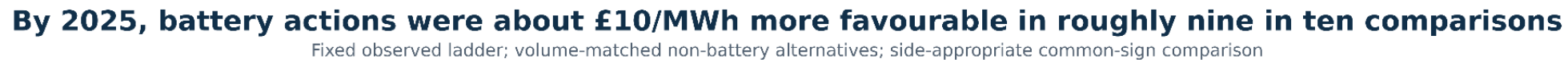

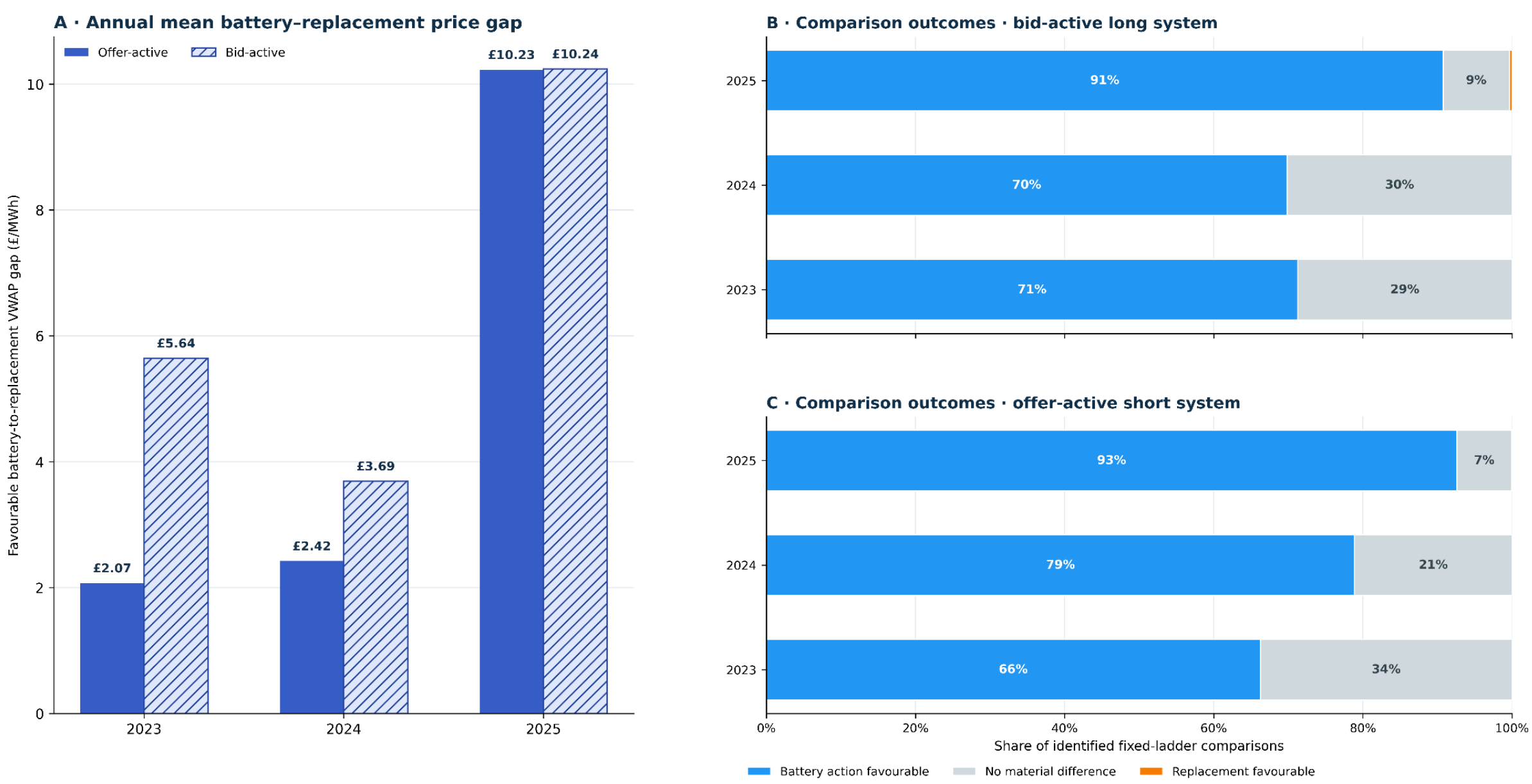

Fig. 6. Fixed-ladder battery action-price advantage relative to volume-matched non-battery alternatives. Panel A reports the annual mean battery-to-replacement VWAP gap by active side; Panels B and C report the distribution of comparison outcomes.

# 5. Discussion

## 5.1 Price-taking, price-making and the position between them

The storage literature generally treats batteries either as price-takers optimising against prices they do not influence, or as strategic price-makers able to preserve spreads through withholding [11,12,13,48,53,67]. The results here show that this binary is incomplete.

At the level of the individual asset, the market structure is not one in which unilateral price-making is possible. Concentration is a necessary condition for unilateral market power, and attributed battery marginality is very far from concentrated. By 2025 it was spread across 137 units with a Herfindahl-Hirschman Index of 0.0151 and a leading-five share of 13.8%; no battery entered the annual top 15 marginal BMUs in any year of the window, and the highest-ranked reached only 23rd. Capacity normalisation adds nuance, since the leading-five comparison reverses once marginal weight is expressed per MW, which indicates that some batteries were relatively intensive marginal participants for their size. Battery HHI nonetheless remained below that of CCGT on the normalised basis in both 2024 and 2025, so the structural conclusion is unchanged.

This is a claim about structure rather than about conduct, and the distinction is worth stating plainly. Marginal frequency cannot by itself identify strategic behaviour, and withholding, the mechanism through which the price-maker literature has storage influencing prices, would if anything reduce the frequency with which a unit is observed at the margin. What the concentration evidence establishes is narrower and firmer: no individual battery occupies a position from which unilateral influence over the balancing price could be exercised, whatever its operator intends. The fixed-ladder comparison bears on behaviour more directly and points the same way, since conditional on reaching the margin, accepted battery actions were priced more favourably than the volume-matched alternatives immediately behind them, which is not the pattern withholding would produce. The sign of that comparison is structural, however, so only its magnitude and trend are informative.

At fleet level, however, storage is now endogenous to routine price formation. Battery-marginal offer prices were broadly comparable with CCGT and modestly lower in two of three annual comparisons, and the fixed-ladder analysis found battery actions approximately £9 to 10/MWh more favourable than the immediate non-battery alternatives by 2025. Batteries therefore occupy an intermediate position that the existing taxonomy does not name: individual units behave in a manner consistent with price-taking, while the fleet collectively shapes routine marginal price formation. Their under-representation in the upper tail bounds rather than negates that role.

## 5.2 The British imbalance-market paradox has moved to the price tail

Atherton et al. [8] describe a paradox in which fossil-fuelled technologies retained disproportionate importance as marginal sellers despite declining fossil generation and rising renewable penetration. At aggregate fuel-type level that dominance has weakened. Applying their four-category taxonomy to the present sample (Fig. 7), the gas share of identified marginality fell by 20.6 percentage points, from 89.7% to 69.1%, while hydro including pumped storage rose by 13.0 points and wind by 10.3 points. Coal fell from 4.2% to 1.5%.

**Under a comparable four-category taxonomy, marginality became substantially less gas-dominated**

Atherton et al. (2017–2022) compared with this study (2023–2025)

90%
80%
70%
20%
15%
10%
5%
0%
89.7%
Gas
-20.6 percentage points
69.1%
18.6%
Hydro (including pumped storage)
+13.0 percentage points
10.8%
Wind
+10.3 percentage points
5.6%
4.2%
Coal
-2.7 percentage points
1.5%
0.5%
Atherton et al.
2017-2022
This study
2023-2025

Battery, excluded from the four-category comparison: 16.3% of the full 2023-2025 sample

Fig. 7. Comparison of marginal-technology shares in Atherton et al. [8] for 2017 to 2022 and this study for 2023 to 2025 under a common four-category taxonomy. Battery storage, which accounts for 16.3% of the full 2023 to 2025 sample, has no counterpart category in the earlier framework.

Resolving marginality by active side and by position within the price distribution shows why the paradox has changed form rather than disappeared. The transition is not one in which VRES directly replace gas at the margin: wind supplied 29.7% of GB generation in 2025 but occupied the price-setting position in only 5% of the sample, and that role was concentrated in long-system bid periods where curtailment bids reflect foregone support payments. Batteries instead emerged as the flexible technology connecting a renewable-rich generation mix to routine real-time balancing and price formation. The principal structural change is therefore the growing role of storage in performing balancing functions previously associated with CCGT on the bid side and pumped storage on the offer side.

Batteries captured a substantial share of routine marginality and overtook CCGT on the bid side in 2025 Q3, while CCGT retained short-system leadership and, together with pumped storage, greater upper-tail prominence. The paradox has therefore shifted: it is weaker in routine marginality but persists in the highest-priced short-system outcomes. Pumped storage illustrates this most sharply, with offer-side share falling from 33% to 10% while remaining the most over-represented technology above the offer-side P95 threshold and recording the highest conditional P95 at £203/MWh. Continued CCGT prominence in the upper tail is consistent with Zakeri et al. [7], who find that gas-fired generation retains disproportionate importance in European price setting as low-carbon penetration rises; the present results refine that conclusion by showing that gas can lose routine marginal frequency while remaining prominent in the highest-priced outcomes.

The comparison also exposes the limits of a fossil-versus-renewable taxonomy. Some 35.6% of marginal-candidate rows fall outside the four categories used in the earlier study, and battery storage alone accounts for 16.3% of the sample. The range of technologies participating in imbalance-price formation has broadened well beyond that framing.

## 5.3 Why batteries occupy the centre and not the tail

The data establish weaker battery presence in the upper tail but not its cause. Three explanations are plausible. First, one- to two-hour batteries may be unable to sustain output through the extended short-system events in which the highest prices arise [12,37]. The capacity-normalised concentration results reinforce this distinction, since adjusting for MW captures differences in power rating but not how long that output can be sustained. Second, battery offers reflect the opportunity cost of discharging now rather than later, which may limit how far submitted prices extend above the expected market distribution. Third, batteries may submit high offers but be accepted earlier and at lower prices, leaving more expensive actions to set the observed extreme.

These explanations imply different future paths. Longer-duration fleets should reduce physical exclusion from the tail, whereas opportunity-cost pricing should vary with expected future spreads and scarcity risk. If acceptance ordering contributes, batteries may still submit high-priced offers even where a higher-priced non-battery action ultimately occupies the attributed margin. Because this study observes accepted energy actions rather than full bidding and dispatch decisions, it cannot distinguish among them. Pumped storage offers a partial test: it lost offer-side share faster than CCGT while retaining the strongest upper-tail position, despite providing storage flexibility on the same side as batteries. That pattern is more consistent with duration than with response speed in severe short-system conditions.

## 5.4 From incumbent displacement to battery-on-battery substitution

Greater battery presence did not bring replacement prices closer to the marginal battery price. Including other batteries in the replacement pool reduced the favourable gap by around 28% to 30% in 2023 but by only 9% to 10% in 2025. The replacement pool became increasingly battery-based without price convergence, which indicates rising battery-on-battery substitution but no clear price compression. This may reflect common opportunity-cost pricing or gaps in the eligible ladder, which the present analysis cannot separate.

The growing battery role raises the question of when the fleet could account for a majority of routine marginality. Holding the observed 2025 marginal share per GW constant in each quarter, a 50% battery marginal share would correspond to approximately 8.7 to 11.1 GW of operational capacity on the bid side and 9.1 to 12.1 GW on the offer side, with median thresholds of 9.6 GW and 10.4 GW. These are conditional scenarios rather than forecasts, but the fleet trajectory places them close at hand. The Clean Power 2030 Action Plan sets a deployment range of 23 to 27 GW of GB battery capacity by 2030 against 4.5 GW installed in 2023 [68], consistent with the system-level requirement identified in the NESO clean power advice [69], and near-term build is tracking that path, with roughly 9 GW projected by the end of 2026 [10]. Relative to the 6.5 GW observed at the end of the study window, the 2030 range implies a further three- to fourfold expansion, so the capacity levels associated here with majority marginality would be passed well before 2030.

That comparison also bounds how far the unit-level capacity association can be extrapolated. The primary specification associates an additional 100 MW of BM-active capacity with 0.547 percentage points of quarterly bid-side marginal share. Most battery BMUs in the panel are rated near 50 MW, so the coefficient implies roughly 0.27 points for a typical unit, and across the 137 units marginal in 2025 that is arithmetically consistent with the observed fleet share of about 36%. Applied mechanically to 23 GW of similarly sized units, however, the same coefficient returns a bid-side share above 100%. The association cannot therefore persist unchanged, and the interesting question for the second half of the decade is which of two adjustments absorbs the difference.

The first is crowding: marginal capture per MW falls as batteries increasingly compete against one another rather than against incumbents, and the price distance between a marginal battery and its nearest substitute compresses towards zero. The fixed-ladder results provide the natural test, and so far they point the other way, since admitting other batteries to the replacement pool reduced the favourable gap by only 9% to 10% in 2025 against 28% to 30% in 2023 even as the pool became more battery-based. The second is saturation: batteries occupy the price-setting position in the large majority of routine periods, and the balancing price becomes principally an expression of storage opportunity cost rather than of fuel cost, the regime anticipated for deeply decarbonised systems in which storage bidding rather than marginal fuel determines prices [56]. Distinguishing between them matters for balancing costs, because the first implies that consumer exposure to storage opportunity cost is disciplined by competition while the second implies that it is not. The same expansion also bears on the upper tail: if the 2030 fleet arrives with longer average duration, the physical exclusion that currently keeps batteries out of extended short-system events should weaken, and the residual incumbent advantage identified in Fig. 5 would narrow.

The implications become more material as the cost of balancing the GB system rises. NESO reports balancing costs of approximately £2.7 billion in 2024/25 and projects that they could peak near £8 billion in 2030 under existing network-delivery assumptions [70]. If batteries continue to occupy a growing share of routine marginal actions, their pricing, duration and competitive behaviour will become increasingly relevant to the cost of system balancing. The favourable £9 to 10/MWh positioning identified here suggests a potentially important role, although the fixed-ladder analysis does not estimate system-level cost savings and should not be read as one.

### 5.5 Limitations

Marginal attribution is reconstructed ex post from the eligible energy-action stack and does not reproduce the full NESO optimisation or every component of the official pricing calculation, so the attributed unit is a price-ranked proxy for the price-setting unit rather than a settlement reconstruction. The fixed-ladder analysis is not a dispatch counterfactual, because nearby actions may differ in location, duration, ramping capability and system value, and because the ordering of acceptances is not observed. Important operational variables, including battery state of charge, availability and ancillary-service commitments, are unobserved, which limits interpretation of why particular units reached the margin. The capacity normalisation adjusts for power rating but not for storage duration. Finally, the fixed-effects estimates describe within-panel associations under a two-way fixed-effects design and should not be read as causal effects of capacity on marginal participation.

## 6. Conclusions

This study separates balancing-price formation into bid- and offer-side regimes, resolves it to individual Balancing Mechanism Units, and establishes batteries as routine price-forming participants in the GB Balancing Mechanism.

Four results follow. First, *marginal price formation recomposed asymmetrically*: battery share rose from 0.8% to 36.2% on the bid side against declining CCGT and wind participation, and from 3.2% to 26.9% on the offer side against declining pumped storage, so pooled fuel-type analysis conceals both the pace and the structure of the change. Second, *marginal participation rose faster than fleet capacity*, and unit-quarter fixed effects associate an additional 100 MW of BM-active capacity with 0.55 and 0.48 percentage points of quarterly bid- and offer-side marginal share, an association that survives separate control for BM-active status. Third, *battery marginality was distributed across an expanding fleet*, although capacity normalisation shows that part of its apparent dispersion relative to CCGT reflects smaller unit sizes; no individual battery displayed the repeated marginality of the leading CCGTs. Fourth, *by 2025 battery actions were approximately £9 to 10/MWh more favourable than immediate volume-matched non-battery alternatives*, a margin that persisted when other batteries were admitted to the replacement pool, while batteries remained under-represented by 12.9 percentage points in the highest-priced 5% of short-system periods.

Taken together, these findings update the British imbalance-market paradox for a battery-rich period. Fossil-fuelled technologies have lost part of their routine marginal role, particularly during long-system periods, but CCGT retains short-system leadership and, with pumped storage, greater prominence in the highest-priced offer periods. The paradox has moved from routine marginality into the price tail. They also show that the price-taker and price-maker categories are not exhaustive. No individual battery holds a share of the margin from which unilateral price-making could be exercised, the leading five accounting for 13.8% of battery marginality across 137 units by 2025, yet the fleet has collectively become endogenous to routine balancing price formation.

Two implications follow for research and market design. Balancing-market analysis should treat batteries as endogenous price-forming participants rather than as passive flexibility resources, which matters for any model that takes the balancing price path as exogenous to storage behaviour. And because the remaining incumbent advantage is concentrated in duration-dependent upper-tail conditions, the pace at which longer-duration storage enters the fleet, rather than power capacity alone, is the variable most likely to determine whether the paradox persists. Extending the unit-level attribution developed here to submitted rather than accepted actions, and to storage duration as an explicit covariate, would allow that question to be tested directly.

## Authorship contribution statement

Robert Dalton: Conceptualization, Methodology, Software, Formal analysis, Data curation, Visualization, Writing (original draft). Aidan O'Sullivan: Conceptualization, Methodology, Supervision, Writing (review and editing).

## Declaration of competing interest

The authors declare that they have no known competing financial interests or personal relationships that could have appeared to influence the work reported in this paper.

## Data availability

The settlement, bid-offer and reference data used in this study are publicly available through the Elexon Insights Solution API and NESO data portal. Processed data and the analysis code are available in the project repository.

## Acknowledgements